\documentclass[amsmath,amssymb,aps]{revtex4} 
\usepackage{graphicx} 
\usepackage{bm} 
\usepackage[usenames]{color} 
\usepackage{listings}

\newcommand{\be}{ 
\begin{equation}
	} 
	\newcommand{\ee}{ 
\end{equation}
} 
\newcommand{\beq}{ 
\begin{eqnarray}
	} 
	\newcommand{\eeq}{ 
\end{eqnarray}
} 

\begin{document}

\title{Dynamics of RNA Translocation through a Nanopore}

\author{Julius B. Lucks$^1$} 
\author{Yariv Kafri$^2$}

\affiliation{ $^1$Lyman Laboratory of Physics, Harvard University, Cambridge, MA 02138\\
$^2$ Department of Physics, Technion, Haifa 32000, Israel.}

\date{\today} 
\begin{abstract}

We present a simplified model of the dynamics of translocation of RNA through a nanopore which only allows the passage of unbound nucleotides. In particular, we consider the disorder averaged translocation dynamics of random, two-component, single-stranded nucleotides, by reducing the dynamics to the motion of a random walker on a one-dimensional free energy landscape of translocation. These translocation landscapes are calculated from the folds of the RNA sequences and the voltage bias applied across the nanopore. We compute these landscapes for 1500 randomly drawn two-letter sequences of length 4000. Simulations of the dynamics on these landscapes display anomalous characteristics, similar to random forcing energy landscapes, where the translocation process proceeds slower than linearly in time for sufficiently small voltage biases across the nanopore, but moves linearly in time at large voltage biases. We argue that our simplified model provides an upper bound to the more realistic translocation dynamics, and thus we expect that all RNA translocation models will exhibit anomalous regimes.

\end{abstract}

\maketitle

\section{Introduction}\label{sec:introduction} 


The translocation of polynucleotides through nanopores has been recently studied extensively, both experimentally and theoretically, due to its relevance to important biological processes \cite{Takyar2005,Namy2006,SimonOster1992} (see below) and emerging sequencing technologies \cite{Akeson1999,LiGolovchenko2003}. The standard experimental setup consists of a nanopore embedded in a thin membrane that separates two buffered solutions (see Figure \ref{fig:model} for a schematic.) Initially the cis-side contains the polynucleotides of interest, and when a voltage is applied across the membrane, the electric field couples to the negatively-charged backbone of the polynucleotide to provide a driving force for translocation into the trans-side

The great interest in this system stems from the small dimensions of the nanopore. The most commonly used nanopore is the $\alpha$-hemolysin ($\alpha$HL) pore derived from \emph{Staphylococcus aureus}, which at its smallest diameter is 1.5 nm \cite{Kasianowicz1996,DeamerBranton2002}. The presence of such a narrow pore hinders the passage of double-stranded polynucleotides which arise naturally when, for example, RNA molecules fold on themselves due to attraction between the bases. Synthetic nanopores, in which the pore is carved out of a synthetic membrane such as silicon nitride, are studied as well \cite{LiGolovchenko2003,FologeaGolovchenko2005} and offer the advantage of a range of pore sizes that can be synthesized for application-specific purposes.

In cases where only single-stranded polymers can pass, nanopore translocation makes an ideal model system suited to studying fundamental biological processes in a controlled fashion. Indeed, biological processes that involve translocation of a polynucleotide across a narrow pore include bacteriophage infection \cite{InamdarPhillips2006}, RNA helicases \cite{Jeang2006}, and most importantly the ribosomal protein synthesis machinery \cite{Wintermeyer2004,BlanchardChu2004,Vanzi2005}. As ribosomes move along messenger RNA (mRNA), any loops that have formed due to folding of the mRNA must be broken in order for protein synthesis to continue \cite{Takyar2005}. It is well known that messenger RNA secondary structure can cause a frameshift in the decoding of the mRNA into protein \cite{Namy2006}, and secondary structures that hinder mRNA translocation through the ribosome are considered to be another facet of the complex gene regulation machinery employed by cells. Experimentally, this has been investigated on only the simplest systems of single hairpins \cite{SauerBudge2003}, while theoretical studies have focused on detailed investigations of specific sequences \cite{GerlandHwa2004,BundschuhGerland2005,Kotsev2006}.  We comment that even without any secondary structure of the mRNA, the dynamics of ribosomes can be rich due to the sequence heterogeneity of the mRNA track and the different chemical fuels used by the ribosome \cite{KafriLubenskyNelson2003}.

In contrast to the rather poor understanding of the translocation of structured polymers, much is known about the translocation of unstructured polynucleotides through nanopores. Several experiments have focused on the translocation of single-stranded DNA (ssDNA) or ssRNA homopolymers which cannot bind on themselves \cite{Bezrukov1994,Kasianowicz1996,Akeson1999,Meller2000,MellerBranton2001,DeamerBranton2002,LiGolovchenko2003,Meller2003,FologeaGolovchenko2005,MatheMeller2005,Butler2006}. When the polymer enters the nanopore, it restricts the flow of the buffer ions in the solution and causes a current blockade. By measuring the duration of the blockade, the experiments can measure the distribution of translocation times. In general, these distributions have been shown to be bimodal as a result of the assymetry in translocating either the 3', or 5' end first \cite{MatheMeller2005,LubenskyNelson1999,Matysiak2006}.

The theoretical work on the translocation of unstructured polymers through nanopores is significant and has focused on the configurational entropic barriers to translocation as a result of the pore confinement \cite{SungPark1996,Muthukumar1999,ChuangKardar2002,KantorKardar2004}, the role of the interaction of individual base pairs with the pore \cite{LubenskyNelson1999,MatheMeller2005}, and other details such as the viscous drag caused by moving the negatively charged polymer and its counterions through solution \cite{LubenskyNelson1999}. There has also been significant effort in various simulations including simplified models \cite{ChuangKardar2002,KantorKardar2004,Luo2006,Muthukumar2006,Matysiak2006} and full-atom simulations \cite{MatheMeller2005}.

Most interestingly, there has been discussion concerning regimes of polymer length and applied voltage where the dynamics of translocation become anomalous \cite{KantorKardar2004,Metzler2003}. By anomalous dynamics we mean that either the velocity, $v$, or the diffusion constant, $D$, or both are not well-defined in a limiting sense \cite{Bouchaud1990}. Operationally we can define velocity with the limit $v = \lim_{t \to \infty} \frac{x(t) - x(0)}{t}$, which is trivial when the position, $x(t)$, is governed by a linear relation with time, $t$, $x(t) = vt$. However, in anomalous dynamics, the position can grow as $t^\mu$ with $\mu<1$, a sublinear powerlaw of time, or even slower which makes $v \to 0$ in this limiting sense. Alternatively, when the position grows as $t^\mu$, the typical time it takes a polymer of length $N$ to cross the pore scales as $N^{1/\mu}$.   It is important to identify likely candidates for anomalous dynamics, not only to appreciate why the dynamics may be so slow, but also to help the interpretation of experimental data \cite{Metzler2003} since an experimentally measured velocity becomes dependent on the time-window used in the measurement \cite{KafriLubenskyNelson2003}.

In this work, we focus on understanding the dynamics of translocation of \emph{structured} polynucleotides such as ssRNA or ssDNA that can bind on themselves in a complementary fashion. In particular, we focus on how the requirement of base pair unbinding at the nanopore affects the dynamics of translocation. To this end, we ignore the configurational entropic barriers, nucleotide-pore interactions, and other details that were important in considering unstructured nucleotides. These simplifications are motivated by the fact that, as we show, the fluctuations in the base pairing energies grow as $N^{1/2}$, where $N$ is the length of the translocating ssRNA, while other barriers typically grow logarithmically \cite{KantorKardar2004}. Furthermore, experimental observation show that unstructured single-stranded nucleotides translocate approximately 100 times faster than hairpins which must be first unzipped to translocate \cite{MatheMeller2004}. Thus any base-pair opening requirements are the rate limiting step for the translocation dynamics and we are safe to focus our attention on them.

We show that while the folded structure of the RNA can be rather complicated, the dynamics of translocation behave very similarly to those of a random walker on a random forcing energy landscape (RFEL). Namely, there exists a regime of voltages where the number of translocated bases, $m$, grows sub-linearly as a function of time with $\overline{\langle m(t) \rangle} \sim t^\mu$ with $\mu<1$, where $\langle \ldots \rangle$ denotes a thermal average while the overline denotes an average over realizations of the RNA sequence. 

This paper is organized as follows: In Section \ref{sec:model_and_energy_landscapes_of_RNA_translocation}, we outline the theoretical study of RNA translocation through nanopores by deriving a  Free Energy Landscape (FEL) in analogy with DNA unzipping. Section \ref{sub:multidimensional_non_equilibrium_translocation_landscapes} discusses important differences between DNA unzipping and RNA translocation in terms of the branched nature of the RNA translocation process. We also discuss a simplifying assumption used in this work to enable us to examine only one-dimensional FELs. Section \ref{sub:calculation_of_translocation_landscapes} describes how these one-dimensional landscapes are calculated for RNA translocation. We present results of the landscapes and model dynamics performed on the landscapes in Section \ref{sec:results}, and conclude with a discussion of how these dynamics can be explained in terms of random forcing energy landscapes, and implications for experiments in Section \ref{sec:discussion}.

\section{Model and Energy Landscapes of RNA Translocation}\label{sec:model_and_energy_landscapes_of_RNA_translocation} 
\begin{figure}
	[htbp] 
	\begin{center}
		(a) 
		\includegraphics[scale=0.8]{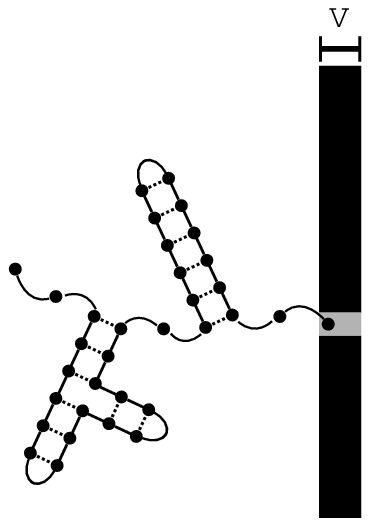} (b) 
		\includegraphics[scale=0.8]{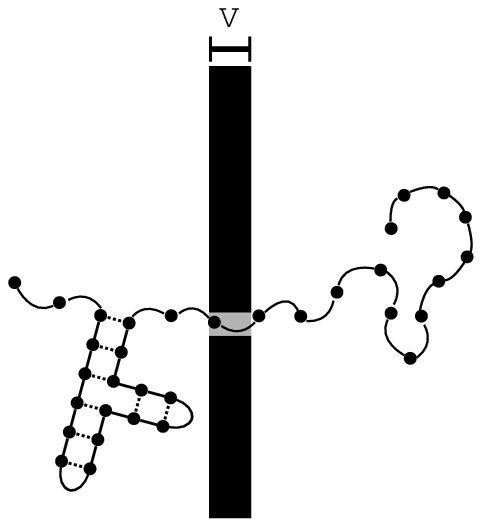} (c) 
		\includegraphics[scale=0.8]{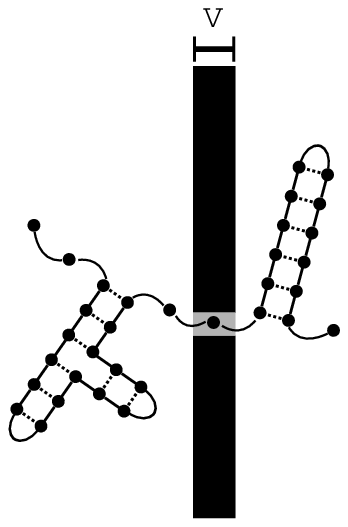} 
	\end{center}
	
	\caption{Model translocation experiment. (a) A negatively-charged single stranded polymer (RNA or ssDNA) starts out on the cis (left) side of a nanopore, which divides a chamber filled with buffer solution. The polymer is a linear string of randomly chosen letters from either a 2 (A,U) or 4 (A,U,C,G) letter alphabet that allows complimentary pairing (dotted lines) (A pairs to U and C pairs to G) and can thus fold on itself (U is replaced by T for DNA). A voltage bias, V, applied across the nanopore (grey) embedded in a membrane (black) provides a driving force for the polymer to translocate to the trans (right) side of the pore where it can be forced to remain unbound to itself (b) (no-refolding), or rebind on itself (c) (refolding). The diameter of the pore only allows single-stranded polymers to pass, and thus all base pairs on the cis-side must be broken in order to translocated.}
\label{fig:model}

\end{figure}

Figure \ref{fig:model} depicts a schematic of a typical translocation experiment. A membrane divides a buffer-filled chamber into cis and trans sides, and initially a single stranded polymer (representing either RNA or equivalently ssDNA) is placed in the cis-side. The RNA molecule has a negatively charged backbone and is composed of a one-dimensional string of randomly chosen nucleotides consisting of the four bases (A,U,C,G) (with U replaced by T for ssDNA). The nucleotides can pair with each other in a complimentary fashion in which A pairs to U (T) and C pairs to G, making the polymer on the cis-side able to fold on itself as depicted in Figure \ref{fig:model}(a) \cite{GerlandHwa2004}. In what follows we refer to a particular matching of the bases as a fold of the RNA molecule.

Embedded in the membrane is a nanopore which is the only pathway between the cis and trans sides. A voltage applied across the nanopore provides the driving force necessary for the negatively-charged polymer to translocate to the trans-side. We assume, similar to the $\alpha$HL nanopore, commonly used in experiments, that the diameter of the pore only allows single-stranded polymers to pass: a base pair that is blocking translocation at the pore in the cis-side must be broken before translocation can proceed, an event which requires overcoming the free energy of the base pair bond (see Figure \ref{fig:model}(a)). If this occurs then the fold of the RNA molecule on the cis-side must readjust to accommodate the loss of a base to the trans-side. 

For the bases that emerge on the trans-side there are two possibilities : (a) they can either be forced to remain unbound or (b) if the specific configuration allows they can rebind creating a fold of the RNA molecule on the trans-side (Figure \ref{fig:model} (b) and (c) respectively.) The former will be referred to from here on as the no-refolding case, and the latter as the refolding case. Both of these cases are examined in detail below. In studying the no-refolding case, we primarily intend to disentangle the complications of refolding on the trans-side from the translocation barriers imposed by the unfolding on the cis-side, rather than mimic an actual experimental setup. However, there are several ways to enforce the no-refolding condition including the addition of single-stranded binding proteins on the trans-side \cite{Ambjornsson2004}, or attaching a bead to the trans end of the RNA, which is pulled by laser tweezers \cite{GerlandHwa2004}. Both methods would add additional complications to our model below, and are not discussed further.

To describe the dynamics of translocation we consider the free-energy change as the RNA molecule is threaded through the pore base-by-base.  As the RNA translocates, the secondary structures changes in order to achieve equilibrium.  The dominant configuration of the RNA in equilibrium, at experimental temperatures, is close to the lowest energy secondary structure, given the constraints imposed by the pore.  In reality, if the change in configuration required to reach the ground state is radical enough, the molecule does not reach this state \cite{KrzakalaMezardMuller2002}.  This makes the problem inherently non-equilibrium in nature, and either requires sophisticated numerics, or simplified models \cite{BundschuhGerland2005}.  In this work, we focus on the latter, with the objective of obtaining a simple picture on which more complicated, realistic models can build.

With this in mind, our model assumes that the system has enough time to reach its (possibly degenerate) ground state at each step of the translocation process. We expect our results to hold even if this assumption is not strictly valid, as long as the RNA configurations are close to the ground state. A similar approach has been applied successfully to the study of translocation of \emph{unstructured} single-stranded polymers through nanopores, where the assumption is validated both through arguments concerning rate-limiting timescales in the problem, and through comparison to experimental results \cite{LubenskyNelson1999,Flomenbom2003,Matysiak2006}. 

Under these assumptions the free energy change as a single base pair is translocated has several contributions, with the most important for \emph{structured} polynucleotides being (a) the free energy gained due to the voltage bias, (b) the free energy required to unbind the base on the cis-side (if it was bound) and (c) the free energy gained due to any rebinding of the base on the trans-side (if possible). There are additional free energy barriers to translocation including entropic barriers due to physical polymer configurations \cite{SungPark1996,Muthukumar1999}. However, under the equilibrium assumption mentioned above, both of these entropic barriers contribute logarithmic terms to the free energy \cite{ChuangKardar2002,KantorKardar2004}, which can be neglected compared to the much larger barriers of breaking and reforming base pairs \cite{GerlandHwa2004} for long enough RNA molecules. These barriers, as we show, give contributions to the free energy that grow as $N^{1/2}$ where $N$ is the length of the translocating molecule.

Even with the simplification of just including base pair breaking and reforming, along with the free energy gained due to the voltage bias, the free energy landscape that describes how the free energy changes as each base is translocated is a highly non-trivial structure. We examine this next, and construct a simplified \emph{one-dimensional} landscape which we use to study the dynamics of the translocation process.

\subsection{Branched, Non-Equilibrium Translocation Landscapes}\label{sub:multidimensional_non_equilibrium_translocation_landscapes} 
\begin{figure}
	[htbp] 
	\begin{center}
		(a) 
		\includegraphics[scale=1]{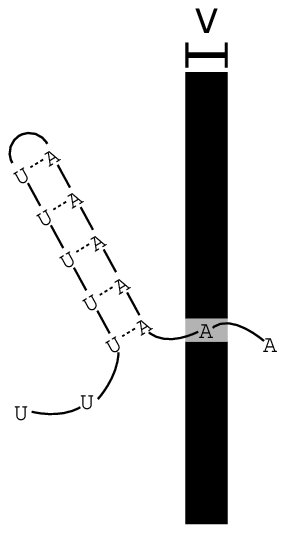} (b) 
		\includegraphics[scale=0.6]{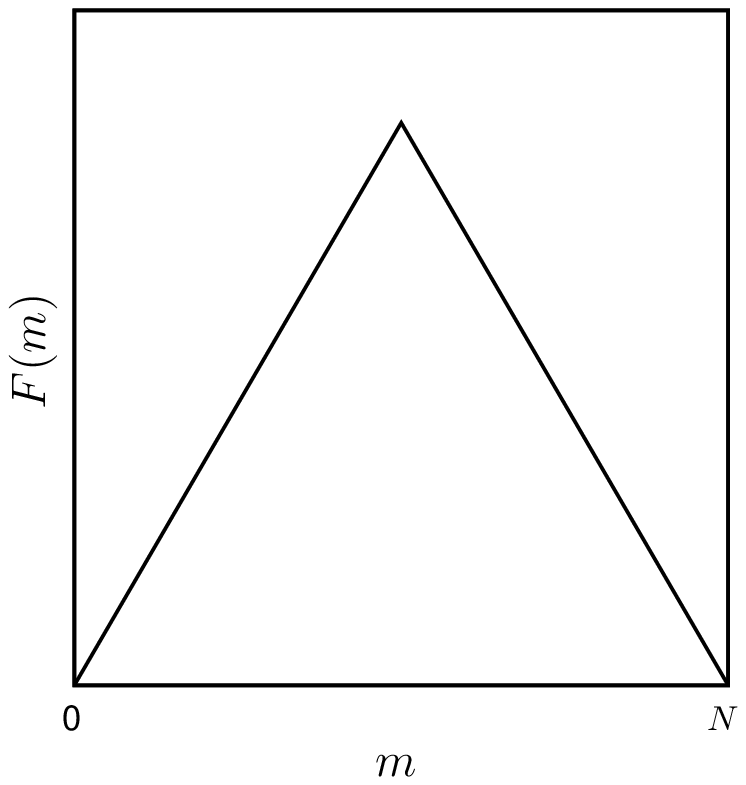} 
	\end{center}
	
	\caption{Schematic of the translocation of an RNA hairpin. The RNA molecule is composed of $N/2$ A's followed by $N/2$ U's, causing all bases to be paired in the initial fold, when the molecule is completely on the cis-side of a membrane (black) with an embedded nanopore (grey). (a) The RNA is shown with one end threaded through the nanopore to the trans-side, after two bases have unbound in order to translocate. (b) Schematic of the Free Energy Landscape for translocation (Eq. \ref{eq:FEL_hairpin}) at the critical voltage such that $F(0) = F(N)$.}
	
	\label{fig:hairpin} 
\end{figure}

Consider first a simple RNA molecule of length $N$ with a hairpin fold in the no-refolding scenario. (Figure \ref{fig:hairpin}(a)). The first $N/2$ bases are $A$ and are complementary to the last $N/2$ $U$ bases, causing all bases to be paired in the initial fold, when the molecule is completely on the cis-side. A sufficient voltage applied across the nanopore provides the driving force necessary to unzip the hairpin to enable translocation \cite{SauerBudge2003,MatheMeller2004}. Since the cooperativity parameter associated with RNA base stacking is small enough to prevent the opening of base pairs at typical experimental temperatures \cite{BlosseyCarlon2003}, we can ignore fluctuations about this ground state hairpin configuration throughout. The problem is then equivalent to unzipping a small double stranded DNA segment by translocating it through a nanopore under applied voltage \cite{Kotsev2006}, or more simply, unzipping double stranded DNA under a constant force \cite{LubenskyNelson2002,Weeks2005}.

As successive base pairs are unzipped, the translocation process steps through a set of well-defined intermediate states labeled by the number of bases that have translocated to the trans side, $m$, each with a well-defined free energy, $F(m)$, where $0 \leq m \leq N$. Consider the change in free energy when the first base is translocated in Figure \ref{fig:hairpin}(a), $F(1) - F(0)$, where we have set the zero of free energy such that $F(0) = 0$. In order for this base to translocate, the system must overcome an energy barrier of $\epsilon$ to break the base pair, but gains an energy $\eta$ from the applied voltage giving $F(1) - F(0) = \epsilon + \eta$. In fact, this is true for $F(m) - F(m-1)$ for $m \leq N/2$ since all of these bases have complementary partners on the cis-side. For $m > N/2$, we have $F(m) - F(m-1) = \eta$, since no base pairs have to be broken. The free energy for state $m$ is thus equal to the sum of all the free energy changes to get there
\begin{equation}
	F(m) = \sum_{i=1}^m \left[ F(i) - F(i-1) \right], 
\end{equation}
or 
\begin{equation}
	\label{eq:FEL_hairpin} F(m) = \left\{ 
	\begin{array}{c}
		m\epsilon + m\eta, \quad m \leq N/2 \\
		(N/2)\epsilon + m\eta , \quad m > N/2 
	\end{array}
	\right. ,
\end{equation}
where the reduced voltage $\eta = \eta(V) = e_{eff} V$, with $e_{eff}$ the effective charge per nucleotide and $V$, the voltage drop across the membrane, which is essentially equal to the applied voltage in experiments. A schematic of Eq. \ref{eq:FEL_hairpin} is shown in Figure \ref{fig:hairpin}(b) at a voltage, $\eta_c$, such that $F(N) = F(0)$. We can see that the necessity to unzip base pairs before translocating gives a significant barrier hindering the process, unless a sufficient voltage is applied. This \emph{one-dimensional} free energy landscape picture leads naturally to a description of the translocation process in terms of a walker moving on the landscape governed by these local energy barriers \cite{Weeks2005}.

This description is however over simplified and even on this level one can see that the RNA translocation problem is more subtle and complicated than the analogy to the DNA unzipping problem in the hairpin example. In the description above we assumed there are no degenerate folds with the same energy on the cis-side. However, consider the state $m=1$, where one base has been translated. There are now $N/2$ U's and $N/2 - 1$ A's on the cis-side, with one $U$ left unpaired. There are $N/2$ choices for this unpaired $U$, thus making this state highly degenerate \cite{DNA_hairpin_degeneracy}. Despite the fact that in the initial $N/2$ translocation steps a base pair always needs to be opened for translocation to occur, the degeneracy changes with $m$.
\begin{figure}
	[htbp] 
	\begin{center}
		\includegraphics[scale=0.45]{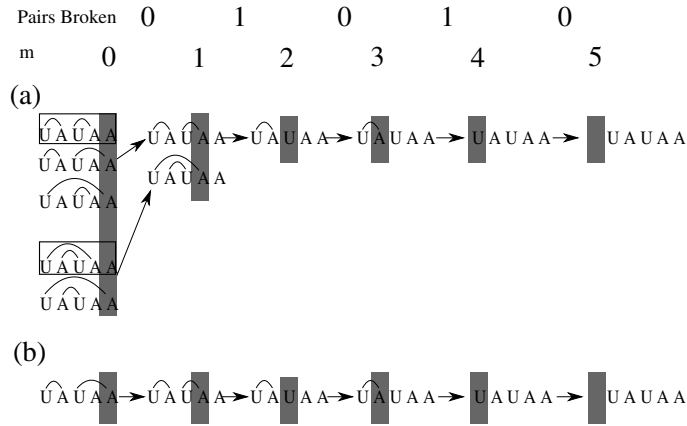} 
	\end{center}
	\caption{Schematic of RNA translocation with no trans-side refolding. The example sequence UAUAA is depicted translocating across a nanopore (grey), which only allows the passage of single nucleotides. Secondary structure bonds between nucleotides are depicted as arcs between letters, and the net bonds broken at each step are displayed as numbers in between translocation steps. The number of bases translocated, $m$, is indicated above each pore. (a) The translocation process is illustrated showing all degenerate folds. Initially the complete sequence is folded on the cis-side in one of five possible folds, each with two base pair (see Figure \ref{fig:nussinov_folding}). When the first nucleotide crosses the pore (first column), three of these structures require a bond breakage, while the other two do not (boxed). These three structures can revert to one of the other two (arrows) by refolding on the cis-side giving a net zero bonds broken in this step of translocation. In the second step of this process, the second nucleotide enters the pore, forcing a bond to be broken in both possible structures. Since the energetics of translocation for all possible structures reduce down to the energetics of the translocation of the topmost structure, in this study we only examine its translocation, giving a well-defined, one-dimensional energy landscape. Note that this corresponds to the quickest possible translocation, since the minimum number of base pairs have to be broken to translocate the sequence. (There are no breakages followed by refoldings). The sum of the number of bonds broken equals two, corresponding to the maximal number of base pairs possible for this structure. (b) The same process depicted in (a) for the chosen sequence.} \label{fig:translocation}

\end{figure}

For random RNA sequences it is well known that the energetic states of RNA folds are highly degenerate \cite{deGennes1968,BundschuhHwa2002} (Appendix \ref{sec:nussinov_folding_algorithm}) and the dynamics can be complicated in a much more severe manner than described in the case of the hairpin. Figure \ref{fig:translocation}(a) depicts a schematic of rna translocation for the sample sequence UAUAA in which no refolding is allowed on the trans-side. Initially this sequence can assume one of five degenerate folds, each with two base pairs on the cis-side \cite{loop_entropy_footnote}. After the first base has been completely translocated there are now only two degenerate states. Some of the initial states require \emph{bond rearangement} (breakage followed by re-pairing) to collapse onto these two states, while the others do not. This makes the description of the process in terms of a one-dimensional energy landscape ill-defined within this energetic model. In fact, the energy landscape that embodies this translocation process is highly branched. At every translocation step, there are branches for each of the possible folds, with appropriate energy barriers.

Describing these energy landscapes is a formidable task and instead of considering the full multidimensional landscape, we project these degenerate states onto an effective one-dimensional landscape. To this end, we describe the translocation process by only considering the \emph{net change} in base pairs between translocation states labeled by the one-dimensional coordinate $m$, representing the number of bases translocated to the trans side. By ignoring all bond rearangements during translocation, we are effectively removing many possible intermediate energy barriers that are present in the full, highly-branched landscape. That is, within the full landscape, we always choose a trajectory with the minimum number of base pairs rearangements (openings and closings). Note that this makes both forward and backward steps of the translocation process happen between well-defined states.

With this choice then, the RNA translocation process can be modeled as the motion of a random walker on a one-dimensional effective free energy landscape, just as in the case of modeling the dynamics of dsDNA unzipping \cite{Weeks2005}. This landscape represents the minimum number of base rearangements that have to be undergone, and thus the dynamics of the random walkers on the landscape are only an \emph{upper bound} to the true dynamics of the system. In what follows below, we systematically investigate the free energy landscapes and dynamics of both the no-refolding and refolding cases for random sequences drawn from the 2 letter model.

\subsection{Calculation of Translocation Landscapes}\label{sub:calculation_of_translocation_landscapes} 

In order to calculate the free energy landscapes of translocation, we need to find the total number of base pairs on both the cis and trans sides at each step of translocation. Since we are assuming the system reaches the ground state at each step, we need to find the optimal fold of each of the subsequences confined to the cis and trans sides for a given $m$. For simplicity, we choose the simplest maximal-matching folding algorithm in order to find these optimal folds. Most notably, we ignore knotted or pseudo-knotted configurations \cite{NussinovJacobson1980}.

The Nussinov folding algorithm \cite{NussinovJacobson1980}, outlined in Appendix \ref{sec:nussinov_folding_algorithm}, inputs a base sequence representing an arbitrary RNA molecule, and returns a matrix $M$, whose elements $M(i,j)$ represent the maximal number of base pairs allowed within the subsequence starting at base $i$ and ending at base $j$. We consider the sequence to be translocating in the direction such that base $N$ is the first to enter the pore. We can thus use this matrix $M$ to calculate the change in the number of base pairs at each step of translocation, and by doing so ignore any internal rearrangements on either side that keep the number of base pairs fixed.

\begin{figure}
	[htbp] 
	\begin{center}
		\includegraphics[scale=0.6]{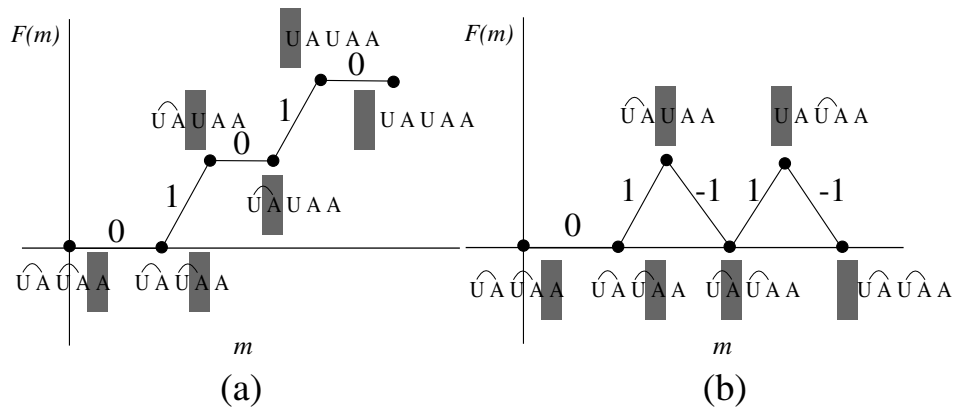} 
	\end{center}
	\caption{Schematic free energy landscapes for translocation, $F(m)$ for (a) no-refolding and (b) refolding for the sample sequence UAUAA. Free energy is in units of $\epsilon$.  Each intermediate state of translocation is labeled with a schematic of the RNA fold (see Figure \ref{fig:translocation}). The landscapes have $\eta = 0$. Between each state is labeled the net base pairs broken: (a) $\sigma(i)$, (b) $(\sigma(i) - \rho(i))$.} \label{fig:translocation_landscape}
\end{figure}

In the case of no-refolding allowed on the trans-side, $M(1,N-m)$ represents the number of base pairs on the cis-side after $m$ bases have translocated. The energetic cost of translocating these $m$ bases is thus the energy required to break a base pair, $\epsilon$, times number of base pairs broken, $M(1,N) - M(1,N-m)$. Each base pair that is translocated also contributes a factor of $\eta$ to the free energy, representing the voltage bias applied across the nanopore. Thus the free energy of the state where $m$ bases have been translocated, $F(m)$ is given by
\begin{equation}
	F_{\mathrm{no-refolding}}(m) = \epsilon (M(1,N) - M(1,N-m)) + \eta m . 
\end{equation}
If we define a variable 
\begin{equation}
	\label{eq:sigma} \sigma(i) = M(1,i) - M(1,i-1) + \eta/\epsilon, 
\end{equation}
this can be rewritten as 
\begin{equation}
	\label{eq:FEL_no_refolding} F_{\mathrm{no-refolding}}(m) = \epsilon \sum_{i=N-m+1}^{N} \sigma(i) 
\end{equation}
The latter form is convenient for considering dynamics on $F(m)$ because of its resemblance to a cumulative sum of random numbers, which is a well-characterized problem (see Section \ref{sec:discussion}) \cite{Bouchaud1990}. Figure \ref{fig:translocation_landscape}(a) depicts the schematic translocation landscape and intermediate states for the sample sequence UAUAA for no-refolding on the trans-side.

In the case of refolding, we must consider how many base pairs can be formed within the subsequence from base $m$ to base $N$, $M(m,N)$. The difference in the number of base pairs formed between translocation steps is embodied in the variable $\rho(i) = M(i,N) - M(i+1,N)$. For simplicity, we consider a pore too thin to accommodate any bases. Since new base pairs on the trans-side contribute a factor of $-\epsilon$ to the free energy, we have
\begin{eqnarray}
	\label{eq:FEL_refolding} F_{\mathrm{refolding}}(m) &= \epsilon \sum_{i=N-m+1}^{N} \sigma(i) - \epsilon M(N-m+1,N) \\
	&= \epsilon \sum_{i=N-m+1}^{N} (\sigma(i) - \rho(i)), 
\end{eqnarray}
where it is understood that $M(N+1,N) = 0$. Figure \ref{fig:translocation_landscape}(b) depicts the translocation landscape and intermediate states for the sample sequence UAUAA for refolding on the trans-side.

Thus for both no-refolding and refolding, the translocation landscapes have the form of a cumulative sum. The dynamics of RNA translocation can be studied in the same way as the dynamics of DNA unzipping by studying the motion of random-walkers on these one-dimensional free energy landscapes \cite{Weeks2005}.

When the cumulative sum is over uncorrelated random variables with a finite variance, the resulting energy landscape is known as a random forcing landscape since the derivative of the free energy is random \cite{Bouchaud1990}. As studies on random forcing energy landscapes show, the dynamics of RNA translocation as governed by FELs from the above equations are anomalous in certain regimes depending on the tilt of the landscape, $\eta$, and the base pairing energy, $\epsilon$. In general the dynamics become anomalous where the overall tilt of the energy landscape is close to being flat. We stress that this is true, provided that the random variables $\sigma(i)$ and $(\sigma(i) - \rho(i))$ are translationally invariant, and independent. 

Below we consider the dynamics on the energy landscapes of RNA translocation that we generate for randomly drawn RNA sequences from the Nussinov folding algorithm. In particular we consider the number of translocated base pairs, $m(t)$, or equivalently the displacement on the energy landscape, averaged over both disorder and thermal realizations. Numerically we perform a thermal average, denoted by $\langle \cdots \rangle$, of the dynamics over single landscapes by repeating the dynamics many times for the same landscape. A disorder average, denoted by  an overline, $\overline{\cdots}$, is obtained from these thermally-averaged dynamics by averaging over many landscapes. We show that in effect the dynamics on these energy landscapes, despite the apparent complexity of the problem, are the same as dynamics on random forcing energy landscapes and exhibit the different regimes of anomalous dynamics, governed by $\epsilon$ and $\eta$, and characterized by the following properties: 
\begin{itemize}
	\item For a specific value of the voltage, where the energy landscape is on average not tilted (see Sec. \ref{sec:results} for a detailed definition) we observe Sinai diffusion where the translocation process proceeds as $\overline{\langle m(t) \rangle} \sim \log^2(t)$
	\item Near the above value of the voltage, the velocity of the transloction process becomes ill-defined and the translocation process proceeds as $\overline{\langle m(t) \rangle} \sim t^{\mu}$, $\mu < 1$. The exponent $\mu$ changes continuously with increasing voltage between $0 < \mu < 1$. 
	\item Finally, outside this regime, for large enough voltages, the velocity becomes well defined with $\mu = 1$, and $\overline{\langle m(t) \rangle} \sim t$.
\end{itemize}

As discussed above, these regimes are an upper bound to the dynamics on the fully branched energy landscapes of RNA translocation.


\section{Results}\label{sec:results}  

In this section, we discuss both the free energy landscapes and dynamics of translocation for the simplified two-letter RNA model system discussed above.  

For a given translocation landscape, dynamics were generated using a standard Monte-Carlo procedure \cite{Weeks2005}.  For each realization, a walker was placed at position $m=0$ of the landscape. At each time step a move $m \to m\pm 1$ was attempted with transition rates
\begin{equation}
	w_{m\to m\pm 1} = \mathrm{min}\{1,e^{-(F(m\pm 1) -F(m))/k_B T}\}.
\end{equation}
with absorbing boundary conditions at the end of the landscape ($m=N$) and reflecting boundary conditions at the beginning ($m=0$). We choose $k_B T=1$ so that that $\epsilon$ and $\eta$ are dimensionless and take $\epsilon = 1$.  The position was then recorded every $S/1000$ steps with $S$ the total number of Monte-Carlo attempts. A thermal average of the position, $\langle m(t) \rangle$, for each realization of the translocation landscape was obtained by repeating the process ten times for each realization, and averaging the position at each of these 1000 time points.

To obtain an average over disorder the process was repeated for 1500 realizations of random two-letter RNA sequences of length $N=4000$.  The disorder averaged dynamics trajectory, $\overline{\langle m(t) \rangle}$, was calculated by averaging $\langle m(t) \rangle$ over realizations of the translocation landscapes. Below we present both the dynamics and the landscapes for a variety of tilts, $\eta$, for both the no-refolding, and refolding cases. We note that changing the value of $N$ did not effect our results.

\subsection{No Refolding}\label{sub:no_refolding}  

Equation (\ref{eq:FEL_no_refolding}) describes the free energy landscape for translocation with no trans-side refolding as a cumulative sum of the variables $\sigma(i)$ (Eq. \ref{eq:sigma}). If $\eta$ is chosen to be sufficiently negative, then the translocation landscape for a particular sequence realization is composed of alternating uphill and downhill segments of varying length (Figure \ref{fig:translocation_landscape}(a)). In our simple model, since we are examining two-letter systems with only one type of base pair, all uphill segments have identical positive slopes, and all downhill segments have identical negative slopes.

\begin{figure}
	[htbp] 
	\begin{center}
		(a) \includegraphics[scale=0.6]{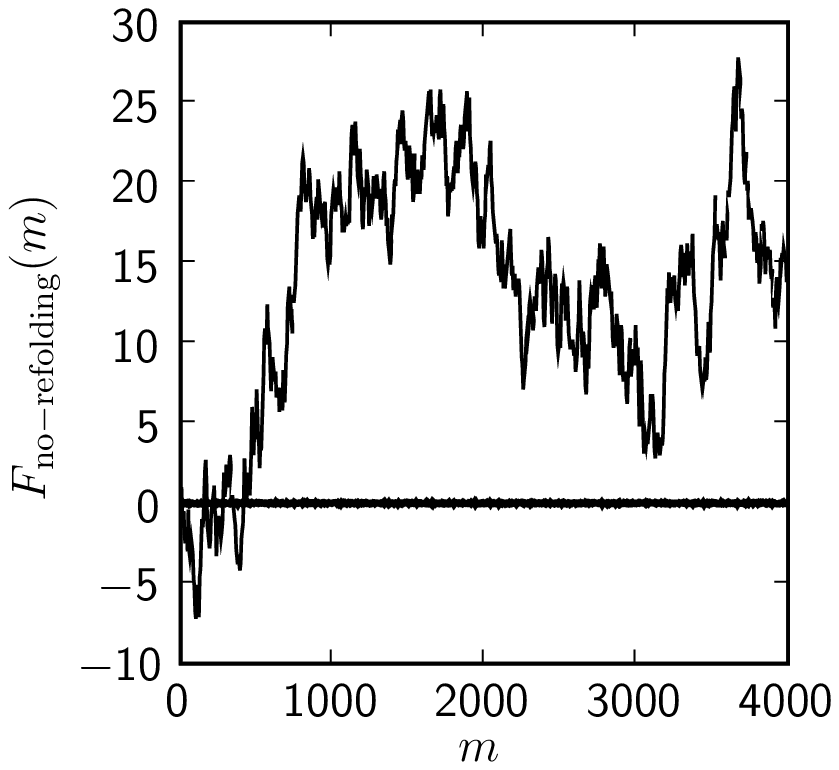} 
		(b) \includegraphics[scale=0.6]{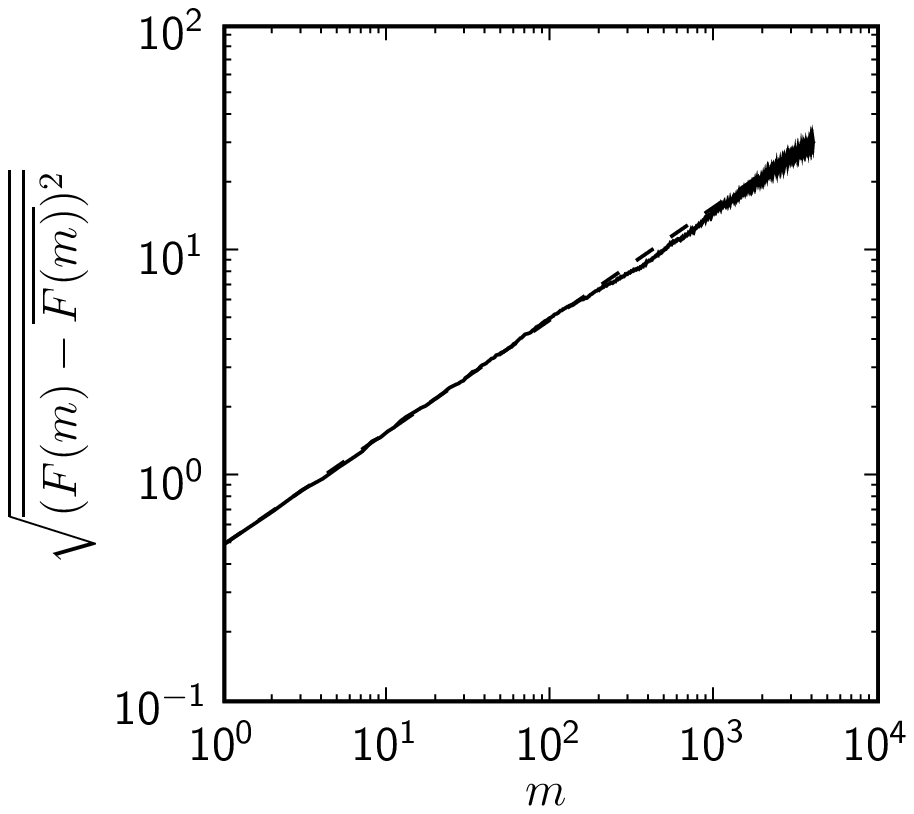} 
	\end{center}
	
\caption{No-refolding translocation landscapes. A sample translocation landscape (a) is shown as a black line, and the average landscape over 1500 realizations, tilted by $\eta_c = -0.5$ is shown on the same figure (straight line).  (b) The standard deviation over disorder $\sqrt{\overline{(F(m) - \overline{F(m)})^2}}$, adjusted by a tilt of -0.006 is shown on a log-log scale with $\sqrt{m}$ shown as a dashed line.}

\label{fig:no_refolding_landscapes}
\end{figure}

Figure \ref{fig:no_refolding_landscapes}(a) shows a sample realization of the no-refolding translocation landscapes at the `critical tilt' of translocation. Similar to random-forcing energy landscapes, the critical tilt is defined as the $\eta$ at which the average over realizations of the landscape has no average tilt. For the 1500 realizations used in this study it is given by $\eta_c = -0.5$. Note that even though the sample landscape does not start and end at an energy of 0, the average over all realizations does. More important for understanding the dynamics are the fluctuations in the energy landscapes which are shown in Figure \ref{fig:no_refolding_landscapes}(b).  These fluctuations, given by $\sqrt{\overline{(F(m) - \overline{F(m)})^2}}$, behave as $\sqrt{m}$ suggesting a structure similar to random forcing energy landscapes, where the fluctuations are known to control the dynamics \cite{Bouchaud1990}.

\begin{figure}
	[htbp] 
	\begin{center}
		\includegraphics[scale=1.0]{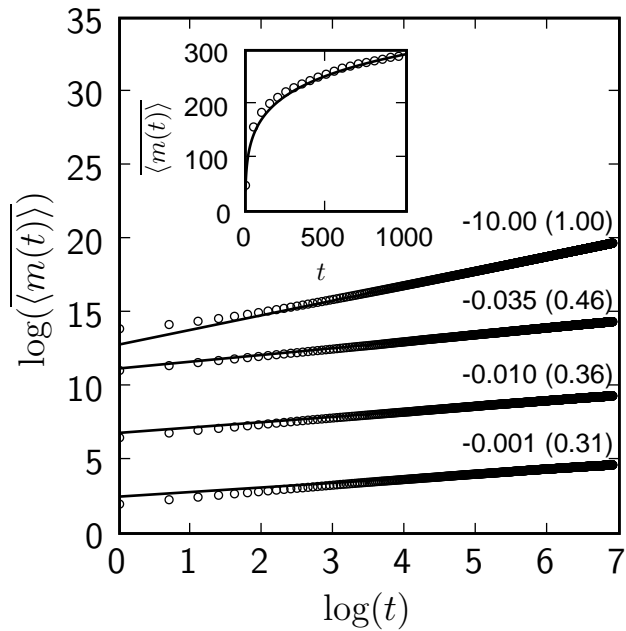} 
	\end{center}
	
	\caption{Trans-side no-refolding, disorder-averaged translocation dynamics.  $\log(\overline{\langle m(t) \rangle})$ vs. $\log(t)$ is plotted  for four different landscape tilts (open circles, shifted for clarity).  The average tilt of the landscape is shown next to each power law fit (solid line), with the exponent of the fit indicated in parentheses.  The total number of steps taken, $S$, was $1\mathrm{x}10^5$ for the three intermediate tilts, and $1\mathrm{x}10^3$ for the high tilt condition.  Inset:  $\overline{\langle m(t) \rangle}$ vs. $t$ for the zero-tilt landscape (open circles) with a fit to $x(t) = A + B\log(Ct)^2$ with $A=29.6$, $B=3.55$, $C=5.5$, and $S=1\mathrm{x}10^7$. Only one out of every 50 points is shown for clarity.}
\label{fig:no_refolding_dynamics} 
\end{figure}

In Figure \ref{fig:no_refolding_dynamics} we show the associated dynamics on these landscapes. Different tilts represent relative tilts to the critical $\eta_c$ listed above. As a quick examination of the energy landscapes suggests, the dynamics are very similar to those exhibited by random walkers on random forcing energy landscapes. Indeed, the dynamics of no-trans-side refolding translocation at the critical tilt follows $\overline{\langle m(t) \rangle} \sim \log^2(t)$, a characteristic of Sinai dynamics. As the tilt is gradually increased, the dynamics obey a sub-linear powerlaw, $\overline{\langle m(t) \rangle} \sim t^{\mu}$, where $\mu < 1$, which changes continuously to $\mu = 1$ for high enough tilts.


\subsection{Refolding}\label{sub:refolding} 

Equation (\ref{eq:FEL_refolding}) describes the free energy landscape for translocation allowing trans-side refolding as a cumulative sum of the variables $\sigma(i)$ (Eq. \ref{eq:sigma}) and $\rho(i)$. If we set $\eta = 0$, then $F(0) = F(N) = 0$ since there will be the same number of base pairs formed in the optimal fold on the cis-side before translocation and the trans-side after complete translocation. Therefore $\eta_c = 0$ represents the `critical tilt' for trans-side folding translocation (Figure \ref{fig:translocation_landscape}(b)). We expect that at this tilt the energy landscape will be peaked on average at $m=N/2$.

\begin{figure}
	[htbp] 
	\begin{center}
		(a) \includegraphics[scale=0.6]{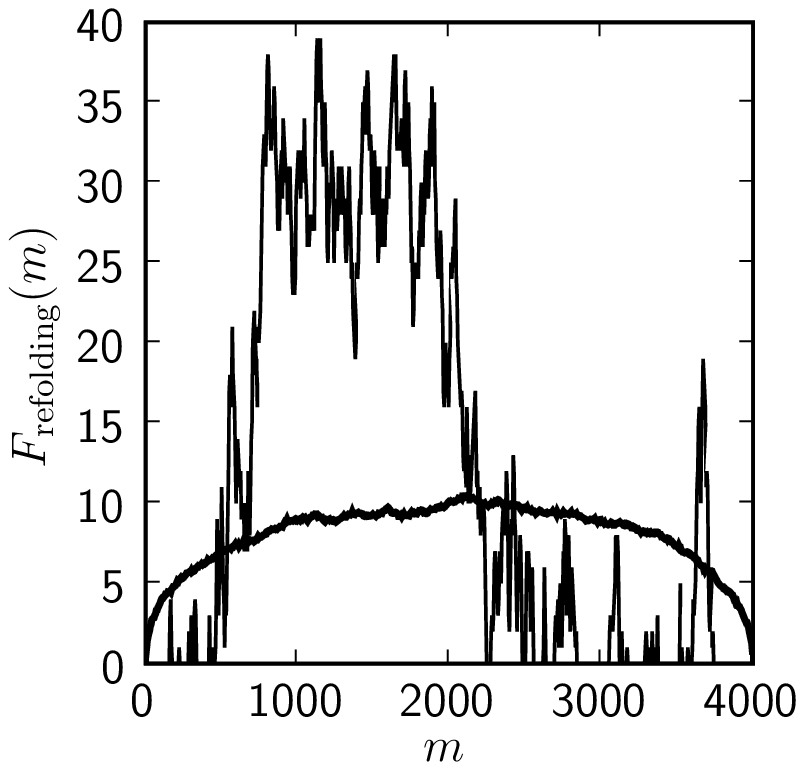} 
		(b) \includegraphics[scale=0.6]{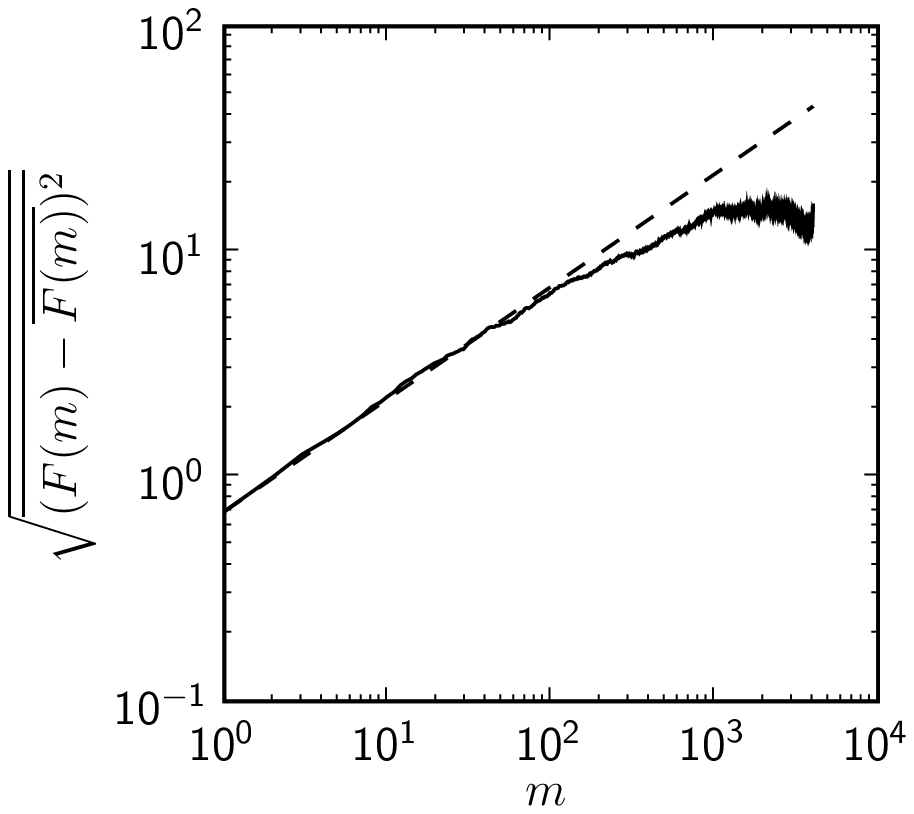} 
	\end{center}
	
    \caption{Refolding translocation landscapes. A sample translocation landscape (a) is shown as a black line, and the average landscape over 1500 realizations, tilted by $\eta_c = 0$ is shown on the same figure.  The sample landscape, as well as the average over all realizations, starts and ends at an energy of 0, since the maximum number of base pairs is the same when the sequence is completely on the cis-, or trans-sides.(b) The standard deviation over disorder $\sqrt{\overline{(F(m) - \overline{F(m)})^2}}$, adjusted by a tilt of -0.004 is shown on a log-log scale with $\sqrt{m}$ shown as a dashed line.}
    
\label{fig:refolding_landscapes}
\end{figure}

Figure \ref{fig:refolding_landscapes}(a) shows a sample realization of the refolding translocation landscapes at the `critical tilt' of translocation. We note that, despite the simplified model, the average landscape is similar to the translocation landscape for a random \emph{four}-letter sequence calculated with a more sophisticated method that fully folds the RNA at each step using experimental free energy parameters (\cite{BundschuhGerland2005} Figure 3(b) inset). As before, our interest is not just the mean energy landscape but also the fluctuations about the mean. In the refolding case these must vanish at $m=N$ and the expected structure might be more elaborate than the no-refolding case. In general the fluctuations in the energy landscape will also not be symmetric about the mean value. In particular, negative fluctuations are less relevant since they are bounded by the minimal value of the energy landscape which is zero. Despite these complications, in Fig. \ref{fig:refolding_landscapes}(b) we see that for a range of $m$, the positive fluctuations grow as $\sqrt{m}$. 

\begin{figure}
	[htbp] 
	\begin{center}		
		\includegraphics[scale=1.0]{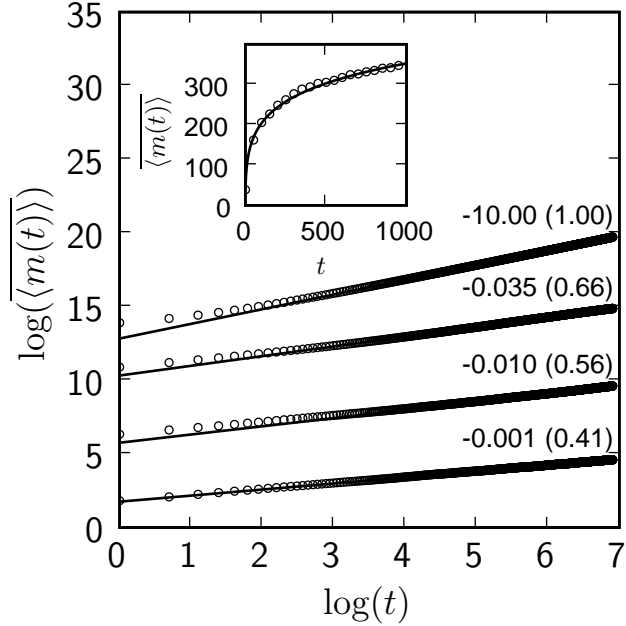}
	\end{center}

    	\caption{Trans-side refolding, disorder-averaged translocation dynamics. $\log(\overline{\langle m(t) \rangle})$ vs. $\log(t)$ is plotted  for four different landscape tilts (open circles, shifted for clarity).  The average tilt of the landscape is shown next to each power law fit (solid line), with the exponent of the fit indicated in parentheses.  The total number of steps taken, $S$, was $1\mathrm{x}10^5$ for the three intermediate tilts, and $1\mathrm{x}10^3$ for the high tilt condition.  Inset:  $\overline{\langle m(t) \rangle}$ vs. $t$ for the zero-tilt landscape (open circles) with a fit to $x(t) = A + B\log(Ct)^2$ with $A=29.6$, $B=4.3$, $C=5.8$, and $S=1\mathrm{x}10^7$. Only one out of every 50 points is shown for clarity.}
    
\label{fig:refolding_dynamics}
\end{figure}

The fluctuations in the energy landscape suggest that, again, the dynamics on the early part of the landscape will be similar to those of a random walker on a random forcing energy landscape. However the dynamics at large values of $m$ might be quicker due to the constraint that $F(N) = 0$ imposed by the refolding scenario.  We stress that even if this is the case, the dominating behavior will be that of the small $m$ when anomalous dynamics are present as long as the region of anomalous dynamics grows with the system size $N$, which we have found to be true (data not shown). 

Figure \ref{fig:refolding_dynamics} shows the associated dynamics on these landscapes, where different tilts represent relative tilts to the critical value $\eta_c = 0$. We again observe the same characteristics of anomalous dynamics in the trans-side refolding translocation dynamics in the Sinai $\overline{\langle m(t) \rangle} \sim log^2(t)$ behavior in the inset. The powerlaw dynamics observed are identical to those in Figure \ref{fig:no_refolding_dynamics}, with only a slight change in exponent for a given average tilt. Since the tilts represent different relative shifts for no-refolding and refolding, this is not surprising. Once again, a large enough tilt removes the anomalous dynamics and causes $\overline{\langle m(t) \rangle} \sim t$.

In summary, for both cases of no-refolding and refolding we observe dynamics which exhibit the same features as a random walker on a random forcing energy landscape, despite the \emph{a priori} complicated structure of the energy landscape. Namely, for small enough tilts the number of translocated bases \emph{does not} grow as a linear function of time.



\section{Discussion}\label{sec:discussion} 

Section \ref{sec:results} shows that the translocation process for both no-refolding, and refolding, exhibit anomalous dynamics for sufficiently small tilts. It is important to emphasize that due to our choice of including only base pair energetics, and our choice of degenerate folds outlined in Section \ref{sub:multidimensional_non_equilibrium_translocation_landscapes}, the simulated translocation dynamics of this model are \emph{upper bounds} on the speed of the dynamics inherent in more sophisticated models of single-stranded polynucleotide translocation. Since anomalous dynamics are slower than $m(t) \sim t$ by definition, we expect anomalous dynamics to be observed even if additional free energy barriers, or the full degeneracy of the possible folds were included.

To get a better understanding of the emergence of anomalous dynamics in this model system, we rewrite Equations (\ref{eq:FEL_no_refolding}) and (\ref{eq:FEL_refolding}) as
\begin{equation}
	\label{eq:FEL_sum_of_random} F(m) = \sum_{i=0}^m \xi(i) 
\end{equation}
where $\xi(i) = \sigma(i)$ for Equation (\ref{eq:FEL_no_refolding}) and $\xi(i) = \sigma(i) - \rho(i)$ for Equation (\ref{eq:FEL_refolding}). The variable $\xi$ is determined from the results of the maximal-pairing Nussinov folding algorithm (Appendix \ref{sec:nussinov_folding_algorithm}) on a random 2-letter sequence. As stated in Sec. \ref{sub:multidimensional_non_equilibrium_translocation_landscapes} if the variables $\xi(i)$ were independently drawn from identical, say Poissonian, distributions, the free energy landscape described by Equation (\ref{eq:FEL_sum_of_random}) would become a sum of random variables. Therefore, the typical fluctuations in the energy landscape would grow as $\sqrt{m}$. Such landscapes are called random forcing energy landscapes (RFEL's) since the derivative of the energy is the random variable. It is well known \cite{Bouchaud1990} that such energy landscapes exhibit anomalous dynamics when the average tilt of the landscape is small. Specifically, dimensional analysis implies that when the quantity $T \Delta F / \mathcal{V}$, is small, anomalous dynamics occur, where $T$ is the temperature, $\Delta F$ is the average tilt of the energy landscape and $\mathcal{V}$ is the variance of the distribution from which $\xi$ is drawn. 

One way to understand the appearance of anomalous dynamics in random forcing energy landscapes is through a simplified model of a walker on a one dimensional lattice of traps \cite{Bouchaud1990}. Within this simplified picture the trap locations mimic the beginning of uphill segments, and the trap dwell times, $\tau$, are a function of the length of the uphill segment following them. In this model, downhill segments and backward motion between traps are ignored. The trapping time in front of the uphill segment behaves as $\tau(l) \sim e^{\kappa l}$, where $l$ is the length of the uphill segment (in dimensionless units such as the lattice spacing) and $\kappa$ is roughly given by $\kappa=(\epsilon - \eta)$. The distribution of uphill lengths, $l$, is expected to behave as $P(l) \sim \exp(-\alpha l)$. It is straightforward to see that under these assumptions the distribution of dwell times behaves as $P(\tau)\sim \tau^{-1-\mu}$ with $\mu=\alpha/ \kappa$. This broad distribution of dwell times leads naturally to the anomalous dynamics when $\mu$ is small enough. In particular, when $\mu<1$ the average dwell time diverges leading to zero velocity. It can be shown through a straightforward analysis \cite{Bouchaud1990} that this leads to a displacement in time which grows as $m(t) \sim t^\mu$. Similar considerations lead, for example, to anomalous diffusion for $\mu<2$. In particular, for sufficiently large values of $\mu$, or equivalently tilts of the energy landscape, the dwell time distribution tails decay fast enough for the velocity to be well defined.

To get some understanding on the origin of the anomalous dynamics in the translocation process of our model we can look for a similar trap structure in the translocation free energy landscapes. In particular, if the distribution of uphill segment lengths across different trap sites is translationaly invariant, uncorrelated, and Poissonian, it will support the above picture which assumes that the uphill segments are drawn from identically distributed random variables. 

To demonstrate that these conditions are indeed satisfied for the 1500 translocation landscapes from the folds of random 2-letter sequences of length $N=4000$ used in this study, we computed the segment length distributions across trap sites for the no-refolding and refolding cases. For no-refolding, we identify a trap site as a start of an uphill segment. These are inevitably followed by a downhill segment. This choice is natural since uphill segments hinder translocation and `trap' the motion, while downhill segments permit rapid translocation. Using this we record the uphill segment lengths for each trap site and study their statistics.

\begin{figure}
	[htbp] 
	\begin{center}

(a) \includegraphics[scale=0.6]{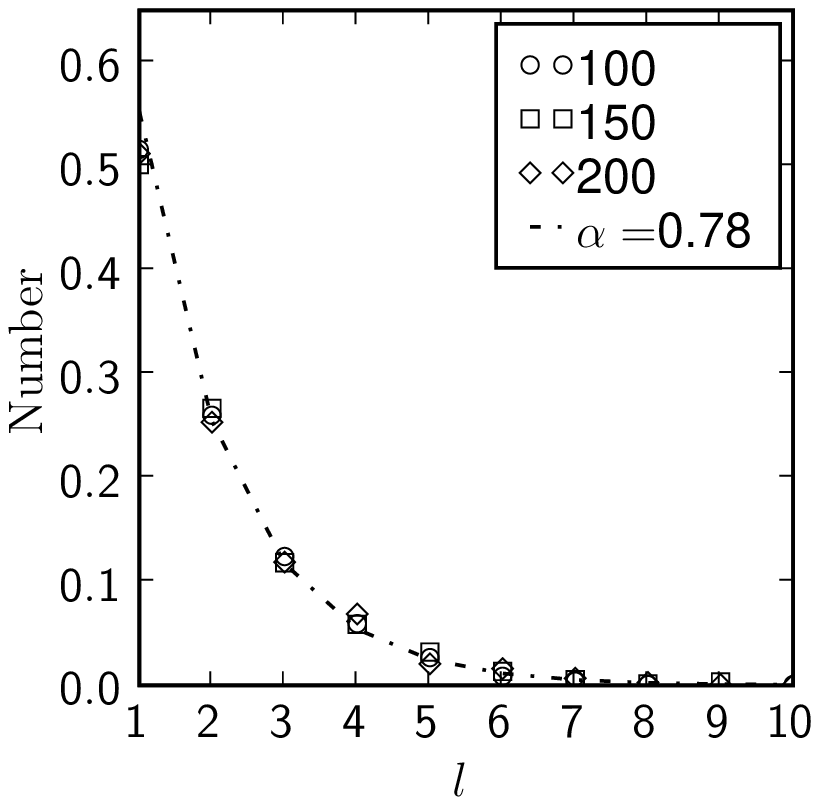} 
(b) \includegraphics[scale=0.6]{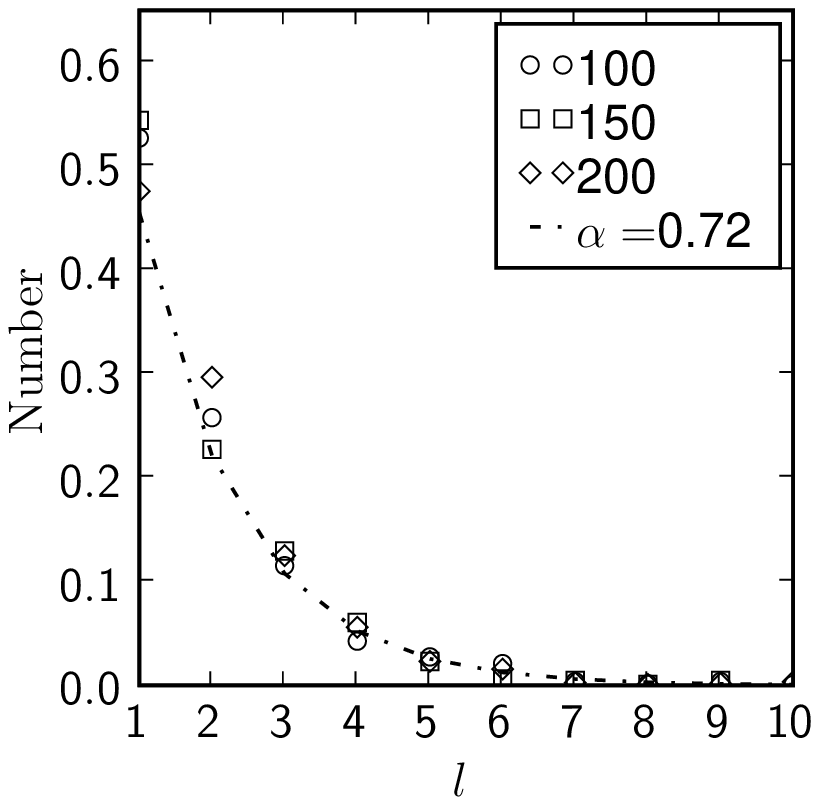} 

	\end{center}
	\caption{Uphill segment length distributions across trap sites for (a) no-refolding, and (b) refolding. For each of the two scenarios, translocation landscapes were partitioned into trap sites by finding the locations of the beginning of uphill segments as discussed in the text. Displayed are normalized histogram heights representing the uphill length distribution for sites numbered 100, 150 and 200 (legend).  These heights were fitted to a Poisson distribution $Ae^{- \alpha l}$, where $l$ is the trap size.}
\label{fig:translational_invariance}
\end{figure}

Figure \ref{fig:translational_invariance}(a) displays the distributions of uphill segment lengths for several trap sites for the no-refolding case. We can see that these distributions are essentially the same across these trap sites, and thus the distribution of uphill lengths across trap sites can be considered translationally invariant. Moreover, the distribution can be fitted very well with a Poisson distribution. We also note that the same analysis on downhill regions of the no-refolding landscapes offer the same conclusions (data not shown).

In the case of refolding the definition of a trap sites is slightly modified, and consists of an uphill-flat-downhill segment triplet. The flat segments lie between an uphill and downhill segment and are due to a compensation event caused by refolding on the trans-side (Section \ref{sub:calculation_of_translocation_landscapes}). If a downhill segment immediately followed an uphill segment, a length of 0 was recorded for that flat segment. Just as in the case of no-refolding, Figure \ref{fig:translational_invariance}(b) indicates that the uphill segment length distribution for refolding landscapes is translationally invariant and Poissonian. The same is true also for the downhill segment length distributions (data not shown).

The presence of flat segments in the translocation landscapes is the primary feature distinguishing the refolding and no-refolding cases. As discussed in Appendix \ref{sec:flat_segments_in_trans_side_refolding_translocation_landscapes}, it is interesting to note that the flat segment lengths become very large, with some translocation landscape realizations composed mainly of one very large flat segment. When a walker encounters a flat segment, it undergoes diffusive motion, which is much faster than the essentially solitary motion of a walker at the base of an uphill barrier. Moreover, with a slight overall tilt, flat segments become downhill regions, while uphill segments remain uphill. Thus refolding landscapes have more effective downhill regions than no-refolding landscapes. Aiming only to get an understanding of the origin of the anomalous dynamics, we do not consider flat segments to contribute to the dwell time at a trap site, and thus leave the discussion of their distribution over trap sites to Appendix \ref{sec:flat_segments_in_trans_side_refolding_translocation_landscapes}. 

\begin{figure}
	[htbp] 
	\begin{center}
		\includegraphics[scale=1]{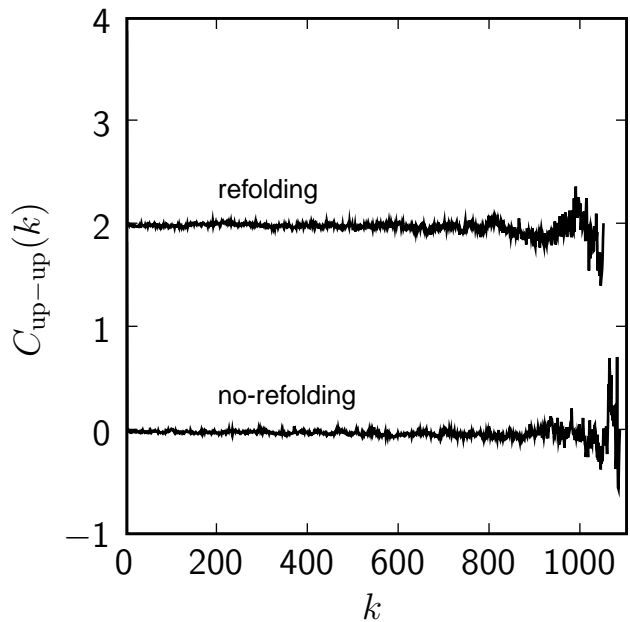} 
	\end{center}
	\caption{Correlation function (Equation \ref{eq:correlation}) for uphill segment lengths for no-refolding and refolding, shifted by 2 for clarity. In both cases there is a markedly pronounced lack of correlations.} \label{fig:correlation}
\end{figure}

To argue that the simple scenario advocated above is plausible we also have to show that there are no long-range correlations between the length of uphill segments at different sites. To do this we compute their correlation across different trap sites, separated by $k$ sites, directly using the equation
\begin{equation}
	\label{eq:correlation} C(k) = \frac{1}{N_T-k}\sum_{i=1}^{N_T-k} (\overline{(u(i)u(i+k))} - \overline{(u(i))} \overline{(u(i+k))}), 
\end{equation}
where $u(i)$ represents the uphill segment length at trap site $i$, $N_T$ is the total number of trap sites, and the overline indicates a disorder average over realizations of the landscapes. Figure \ref{fig:correlation} shows that there are no significant correlations in the uphill segment lengths across trap sites for both no-refolding and refolding scenarios, with similar conclusions for the correlation among downhill segments (data not shown).

Thus the numerics verify that the uphill segment lengths are independently drawn from identically distributed Poisson distributions making the picture presented above for the emergence of anomalous dynamics plausible. The anomalous dynamics observed in Section \ref{sec:results} are thus due to the random forcing nature of the translocation process. In particular, the no-refolding free energy landscapes seem identical to random forcing energy landscapes, despite the complicated fold of the RNA.

We have demonstrated that the dynamics associated with the translocation of RNA sequences in our model are anomalous, but we have made many simplifications. As pointed out in Section \ref{sub:multidimensional_non_equilibrium_translocation_landscapes}, we are ignoring any polymer configurational entropy barriers that might be present such as loop entropies. In addition, we have not included the excluded volume effects of realistic polymer chains. Since the anomalous dynamics we consider here are translocation-rate limiting, we expect that they would not change if these details were included. Furthermore, Figures \ref{fig:no_refolding_landscapes}(b) and \ref{fig:refolding_landscapes}(b) both show that the free energy fluctuations due to the base pairing effects alone, grow as $\sqrt{m}$, which dominates any logarithmic terms introduced by these other effects as mentioned above. We have also ignored any interactions the nucleotides can have with the pore itself \cite{LubenskyNelson1999}. These interactions would modify the $\sigma(i)$ variable in Equation (\ref{eq:sigma}) through a now sequence-dependent $\eta(i)$. It is straightforward to see that these {\it bounded} contributions to the energy landscape will not effect our results.

Finally we have only considered 2-letter systems, but more realistically, we can consider 4-letter systems that are closer to actual RNAs. The possible folds are more complex, but we do not expect that this would remove the anomalous behavior of the translocation dynamics. In fact, due to the similarity between the average refolding translocation landscape in Figure \ref{fig:refolding_landscapes}(a), and the translocation landscape for a random 4-letter sequence studied elsewhere (\cite{BundschuhGerland2005} Figure 3(b) inset), we expect our results to hold also in those systems.


\section{Conclusion}\label{sec:conclusion}  

We have presented a simple model for the translocation of RNA through a nanopore which only allows passage of single-stranded nucleotides. At the heart of this model is the translocation free energy landscape which is calculated from the fold of a particular RNA sequence and the voltage bias applied across the nanopore.

The translocation dynamics, modeled as the motion of a random walker on these calculated translocation landscapes, was shown to display anomalous characteristics such as Sinai and sub-linear power-law regimes for sufficiently small voltage biases across the nanopore. This is despite the complicated nature of the RNA folds. In fact, this might suggest that even more sophisticated models which consider the full RNA folding dynamics should display the same anomalous characteristics.

\section{Acknowledments}\label{sec:acknowledments} 

We are grateful to David Nelson for introducing us to this problem, and for many useful discussion and comments. We also thank Herv\'{e} Isambert, Oskar Hallatscheck and C. Brian Roland for useful discussions, and the Physico-chemie section of L'Institute Curie, Paris, for accommodating us during part of this work. YK thanks the Boston University visitors program for their hospitality. JBL acknowledges the financial support of the John and Fannie Hertz Foundation. This work was supported by the United States-Israel Science Foundation (BSF) and the Israeli Science Foundation (ISF).


\appendix

\section{Nussinov Folding Algorithm}\label{sec:nussinov_folding_algorithm} 

In this section we discuss a modified version of the maximal folding algorithm of Nussinov and Jacobsen \cite{NussinovJacobson1980} for folding two-letter random sequences. We strongly follow the discussion in \cite{NussinovJacobson1980}, but add focus to the topic of degeneracies in the possible secondary structures since it is of importance to the study of RNA translocation.

The Nussinov maximal matching algorithm returns the maximum number of base pairs possible in the secondary structure of a given one-dimensional sequence of complimentary pairing bases. In this work we focus on two-letter sequences, for example, only containing the bases A and U, where A can pair with U, but not with itself and vice versa. At the heart of the algorithm is the $M$-matrix, where $M(i,j)$ is equal to the maximum number of base pairs possible for the subsequence from index $i$ to $j$ inclusive. If $N$ is the length of the sequence being considered, then $M(1,N)$ is the maximum number of base pairs possible in the secondary structure of the whole sequence.

The algorithm starts off with all diagonal entries of $M$ set to 0, since bases cannot pair on themselves. It next considers all possible base pairs between adjacent bases (subsequences of length 2) to fill in the first off diagonal of $M$, setting elements to 1 if the bases can pair on each other, or to 0 if they cannot. The algorithm continues to sweep over all possible contiguous subsequences of increasing length, always using information about maximal pairing in the previous smaller subsequences to determine the maximal pairing in the current subsequence, and filling the entries in $M$.

A code listing \cite{python} with the details of the algorithm is below. The algorithm initializes the $M$-matrix to be an $N\mathrm{x}N$ zero matrix (A), where we are using 1-based subscripting. The outer iteration is over all possible subsequence lengths \verb=2..N=, where \verb=1..3= represents \verb=1,2,3= in succession (B). Consider the subsequence from index \verb=start= to index \verb=end=. For this subsequence, the algorithm iterates over all indices (\verb=intermediate=), and tests whether or not those bases can pair with base \verb=end= (C). A vector (\verb=pair_counts=) stores the number of base pairs that are formed when base \verb=intermediate= pairs with base \verb=end=, which makes use of $M$-matrix elements for the maximal number of bases possible in the subsequences from \verb=start= to \verb=intermediate-1= and \verb=intermediate+1= to \verb=end-1=, since these subsequences are not interrupted by the bond between bases \verb=intermediate= and \verb=end= (D). Once all possible pairings with base \verb=end= are considered, and the no-pairing condition is taken into account, \verb=M(start,end)= is set to the maximum of these stored results (E).

\begin{verbatim}
N = length(sequence)
M = zeros(N,N) #A
for sub_length in [2..N]: #B
    for start in [1..N-sub_length+1]:
        end = start+sub_length-1
        sub_sequence = sequence[start:end]
        N_sub_sequence = len(sub_sequence)
        pair_counts = zeros(N_sub_sequence,1)

        for intermediate = [1..N_sub_sequence-1]: #C
            if sub_sequence[intermediate] == sub_sequence[end]:
                pair_counts = M[start,start+intermediate-1] + 1 + \
                              M[intermediate+1,end-1] #D
        pair_counts[N_sub_sequence] = M[start,end-1]
        M[start,end] = max(pair_counts) #E
return M 
\end{verbatim}

\begin{figure}
	[htbp] 
	\begin{center}
		\includegraphics[scale=0.6]{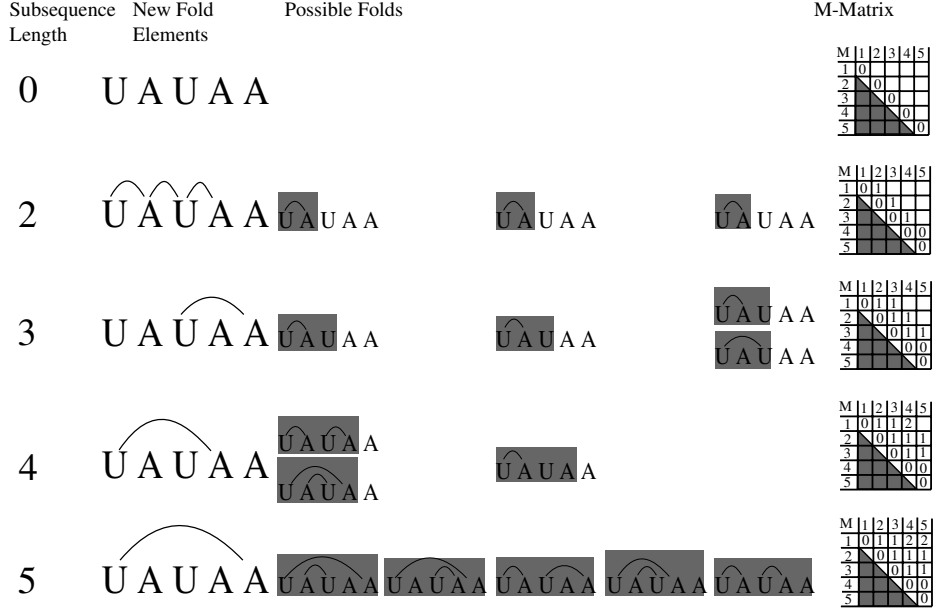} 
	\end{center}
	\caption{Schematic of the intermediate states in the Nussinov folding algorithm applied to the example sequence UAUAA. Successive rows represent increasing values of sub\_length, the subsequence length being considered. The second column indicates new secondary structures that can be formed at this subsequence length, with base pairings denoted by arcs. The next three columns represent the folds including these new elements with grey boxes indicating the subsequence that is folded. Degeneracies are indicated with stacked folds sharing the same subsequence box, except in the last row where these are shown horizontally. At each step, the M-matrix is filled in (last column), and since $M(i,j) = M(j,i)$ only the values for the top half are shown.} \label{fig:nussinov_folding}
\end{figure}

Figure (\ref{fig:nussinov_folding}) depicts a schematic of this algorithm for the sample sequence UAUAA. Once the diagonal entries of $M$ are set to 0, the algorithm considers the folding of all subsequences of length 2 (second row). There are a total of three base pairs possible, setting three values of $M$ to 1. The last base pair between AA is not possible making $M(4,5)=0$. When the subsequence length of 3 is considered, only one new base pairing is possible in addition to the ones obtained through subsequences of length two. This sets $M(3,5)=1$. Note that even though this new element is compatible with the first UA being paired together, this is not counted in $M$ until the full sequence of length 5 is considered (bottom row, middle fold). The algorithm continues in this manner, considering ever larger sebsequences until finally the full sequence is considered.

Note that $M$ stores only the maximal number of base pairs possible, not which folds are actually present. Thus $M(1,5)=2$, denoting 2 base pairs, yet this state is 5-fold degenerate as can be seen in the bottom row. The original Nussinov algorithm has a prescription for recovering the representative secondary structures for these maximal folds, but does not highlight the role of degeneracy in these folds. In this work, this degeneracy is effectively removed by the procedure by which the $M$-matrix is used to calculate the energy landscapes for translocation (Section \ref{sub:calculation_of_translocation_landscapes}).

\section{Flat Segments in Trans-side Refolding Translocation Landscapes}\label{sec:flat_segments_in_trans_side_refolding_translocation_landscapes}  

\begin{figure}[htbp]
    \begin{center}
        (a) \includegraphics[scale=0.6]{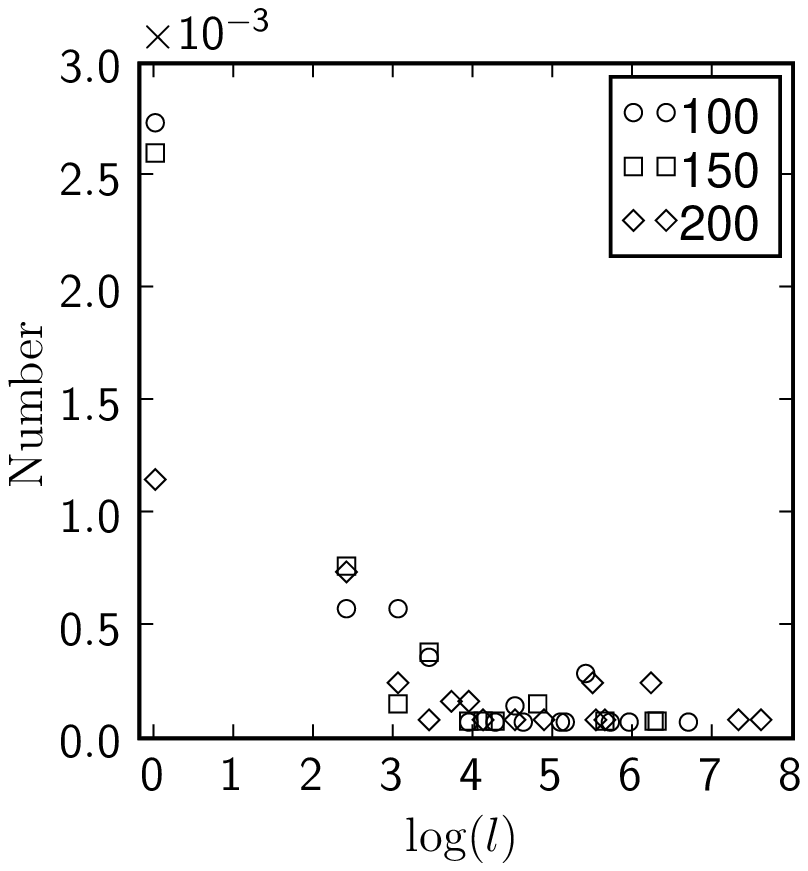}
        (b) \includegraphics[scale=0.6]{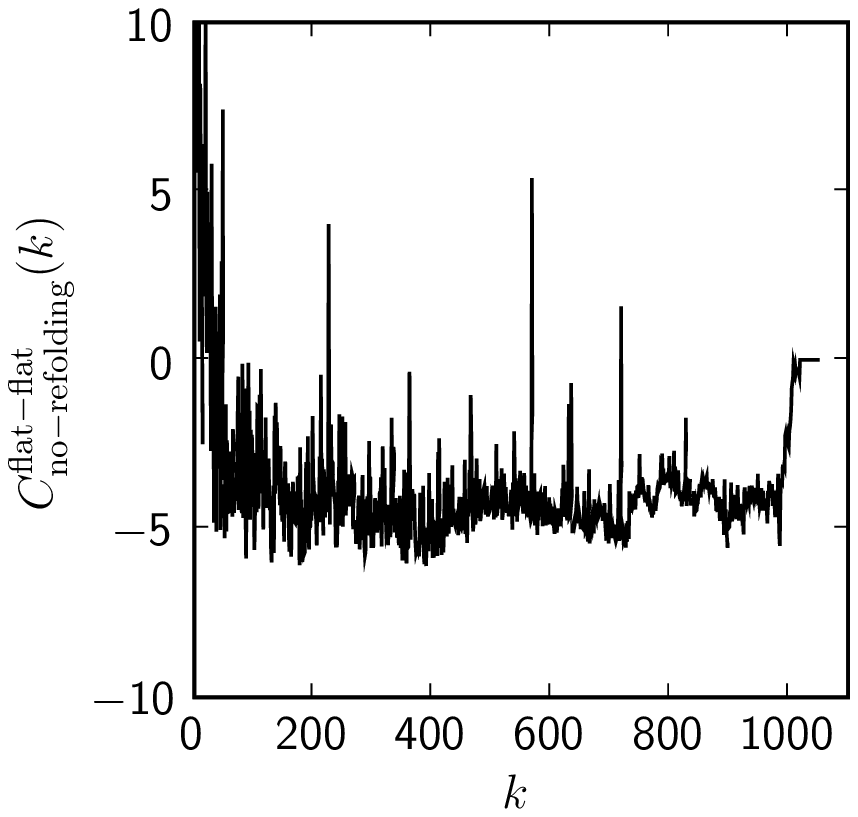}
    \end{center}
    \caption{Refolding flat segment length distribution (a), and correlation function (b).  In (a), just as in Figure \ref{fig:translational_invariance}(b), the flat segment distribution is plotted for the same sites, but versus $\log(l)$.  In (b), the flat-flat correlation function (Equation \ref{eq:correlation}) is plotted.}
    \label{fig:refolding_flat_TI_corr}
\end{figure}

In this section we present the distribution and correlation function for flat segments of the refolding translocation landscapes.  As seen from Figure \ref{fig:refolding_flat_TI_corr}(a), the distribution of flat segments is roughly translationally invariant across different sites, as with the up segment distribution shown in Figure \ref{fig:translational_invariance}(b).  We note however, that this distribution is much more complicated than the up segment distribution, and is much broader as reflected by the $\log(l)$ scale.

The flat-flat correlation function, shown in Figure \ref{fig:refolding_flat_TI_corr}(b), is negative for large length scales.
A given translocation landscape has a finite length, which is equal to the length of the RNA molecule, $N$.  We see from the distribution of flat segment lengths, that these lengths can approach 2000, which is half of the allowed $N$ for a landscape.  We thus interpret the negative correlations as being a result of these two effects: if a large flat segment occurs, there is not room in the translocation landscape for another one of the same length to occur, so small ones occur leading to negative correlations.

However, as discussed in Section \ref{sec:discussion}, these properties of the flat segments do not influence our interpretation of refolding translocation landscapes as random forcing energy landscapes, and our conclusion that this results in anomalous dynamical regimes.

\bibliography{rnat}

\begin{thebibliography}{49}
\bibitem[{DNA()}]{DNA_hairpin_degeneracy}
\bibinfo{note}{Contrast this to the unzipping of a DNA hairpin under constant
  force where the tension on the strands leaves only one state indexed by how
  many base pairs have been unzipped to that point.}
\end{thebibliography}

\end{document}